\documentclass[letter]{aa} 

\newcommand{\mgas}{$M_{\rm gas}$}

\newcommand{\md}{$M_{\rm dust}$ }

\newcommand{\fgas}{$f_{\rm gas}$}
\newcommand{\fdust}{$f_{\rm dust}$}

\usepackage{float}
\usepackage{graphicx}
\usepackage{txfonts}
\usepackage{xcolor}
\usepackage{lipsum}
\usepackage{amsmath}
\usepackage[normalem]{ulem}
\usepackage[para]{threeparttable}
\usepackage{xcolor}
\usepackage{amssymb}

\definecolor{xlinkcolor}{cmyk}{1,1,0,0}
\newcommand{\cooz}{CO \textit{J}=1$\rightarrow$0}
\newcommand{\coto}{CO \textit{J}=2$\rightarrow$1}
\newcommand{\lco}{$L^{\prime}_{\rm CO(1-0)}$}
\newcommand{\lcot}{$L^{\prime}_{\rm CO(2-1)}$}

\newcommand{\hst}{\textit{HST}}

\PassOptionsToPackage{hyphens}{url}
 \usepackage[bookmarks=true,         
     pdfnewwindow=true,      
     colorlinks=true,    
     linkcolor=xlinkcolor,     
     citecolor=xlinkcolor,     
     filecolor=xlinkcolor,  
     urlcolor=xlinkcolor,      
     final=true,
 ]{hyperref}
\begin{document}

   \title{"Dust Giant": Extended and Clumpy Star-Formation in a Massive Dusty Galaxy at $z=1.38$}

   \author{Vasily Kokorev \inst{\ref{aff:kapteyn}} \fnmsep \inst{\ref{aff:dawn}} \fnmsep\thanks{Corr. author, \email{vasily.kokorev.astro@gmail.com}}
          \and
          Shuowen Jin \inst{\ref{aff:dawn}} \fnmsep \inst{\ref{aff:dtu}}
          \and 
          Carlos G\'omez-Guijarro \inst{\ref{aff:saclay}}
          \and
          Georgios E. Magdis \inst{\ref{aff:dawn}} \fnmsep \inst{\ref{aff:dtu}}
          \and
          Francesco Valentino \inst{\ref{aff:eso}} \fnmsep \inst{\ref{aff:dawn}} \fnmsep 
          \and 
          Minju M. Lee \inst{\ref{aff:dawn}} \fnmsep \inst{\ref{aff:dtu}}
          \and
          Emanuele Daddi \inst{\ref{aff:saclay}}
          \and
          Daizhong Liu \inst{\ref{aff:maxplanck}}
          \and
          Mark T. Sargent \inst{\ref{aff:ISSI}}
          \fnmsep \inst{\ref{aff:sussex}}
          \and
          Maxime Trebitsch \inst{\ref{aff:kapteyn}}
          \and
          John~R.~Weaver \inst{\ref{aff:amherst}}
          }

   \institute{Kapteyn Astronomical Institute, University of Groningen, P.O. Box 800, 9700AV Groningen, The Netherlands
  \label{aff:kapteyn}
  \and
            Cosmic Dawn Center (DAWN), Jagtvej 128, DK2200 Copenhagen N, Denmark \label{aff:dawn}
            \and
            DTU-Space, Technical University of Denmark, Elektrovej 327, DK2800 Kgs. Lyngby, Denmark
            \label{aff:dtu}
            \and 
            Universit{\'e} Paris-Saclay, Universit{\'e} Paris Cit{\'e}, CEA, CNRS, AIM, 91191, Gif-sur-Yvette, France
            \label{aff:saclay}
            \and
            European Southern Observatory, Karl-Schwarzschild-Str. 2, D-85748 Garching bei Munchen, Germany
            \label{aff:eso}
            \and
            Max-Planck-Institut fur Extraterrestrische Physik (MPE), Giessenbachstr. 1, D-85748 Garching, Germany \label{aff:maxplanck}
            \and
            International Space Science Institute (ISSI), Hallerstrasse 6, CH-3012 Bern, Switzerland \label{aff:ISSI}
            \and
            Astronomy Centre, Department of Physics and Astronomy, University of Sussex, Brighton BN1 9QH, UK \label{aff:sussex}
            \and
            Department of Astronomy, University of Massachusetts, Amherst, MA 01003, USA \label{aff:amherst}
            }

   \date{Received XXX; accepted XXX}
 
  \abstract 
   {We present NOEMA CO (2-1) line and ALMA 870 $\mu$m continuum observations of a main-sequence galaxy at $z=1.38$. The galaxy was initially selected as a "gas-giant", based on the gas mass derived from sub-mm continuum (log$(M_{\rm gas}/M_{\odot})=11.20\pm0.20$), however the gas mass derived from CO (2-1) luminosity brings down the gas mass to a value consistent with typical star-forming galaxies at that redshift (log$(M_{\rm gas}/M_{\odot})=10.84\pm0.03$). Despite that the dust-to-stellar mass ratio remains elevated above the scaling relations by a factor of 5. We explore the potential physical picture and consider an underestimated stellar mass and optically thick dust as possible causes. Based on the updated gas-to-stellar mass ratio we rule out the former, and while the latter can contribute to the dust mass overestimate it is still not sufficient to explain the observed physical picture. Instead, possible explanations include enhanced HI reservoirs, CO-dark H$_2$ gas, an unusually high metallicity, or the presence of an optically dark, dusty contaminant. Using the ALMA data at 870 $\mu$m coupled with $HST$/ACS imaging, we find extended morphology in dust continuum and clumpy star-formation in rest-frame UV in this galaxy, and a tentative $\sim 10$ kpc dusty arm is found bridging the galaxy center and a clump in F814W image. The galaxy shows levels of dust obscuration similar to the so-called $HST$-dark galaxies at higher redshifts, and would fall into the optically faint/dark $JWST$ color-color selection at $z>2$. It is therefore possible that our object could serve as low-$z$ analog of the $HST$-dark populations. This galaxy serves as a caveat to the gas masses based on the continuum alone, with a larger sample required to unveil the full picture.}

\titlerunning{Dust Giant}
\authorrunning{Kokorev et al.}

   \maketitle
%
 
\section{Introduction}

\begin{figure*}
\centering
\includegraphics[width=0.9\textwidth]{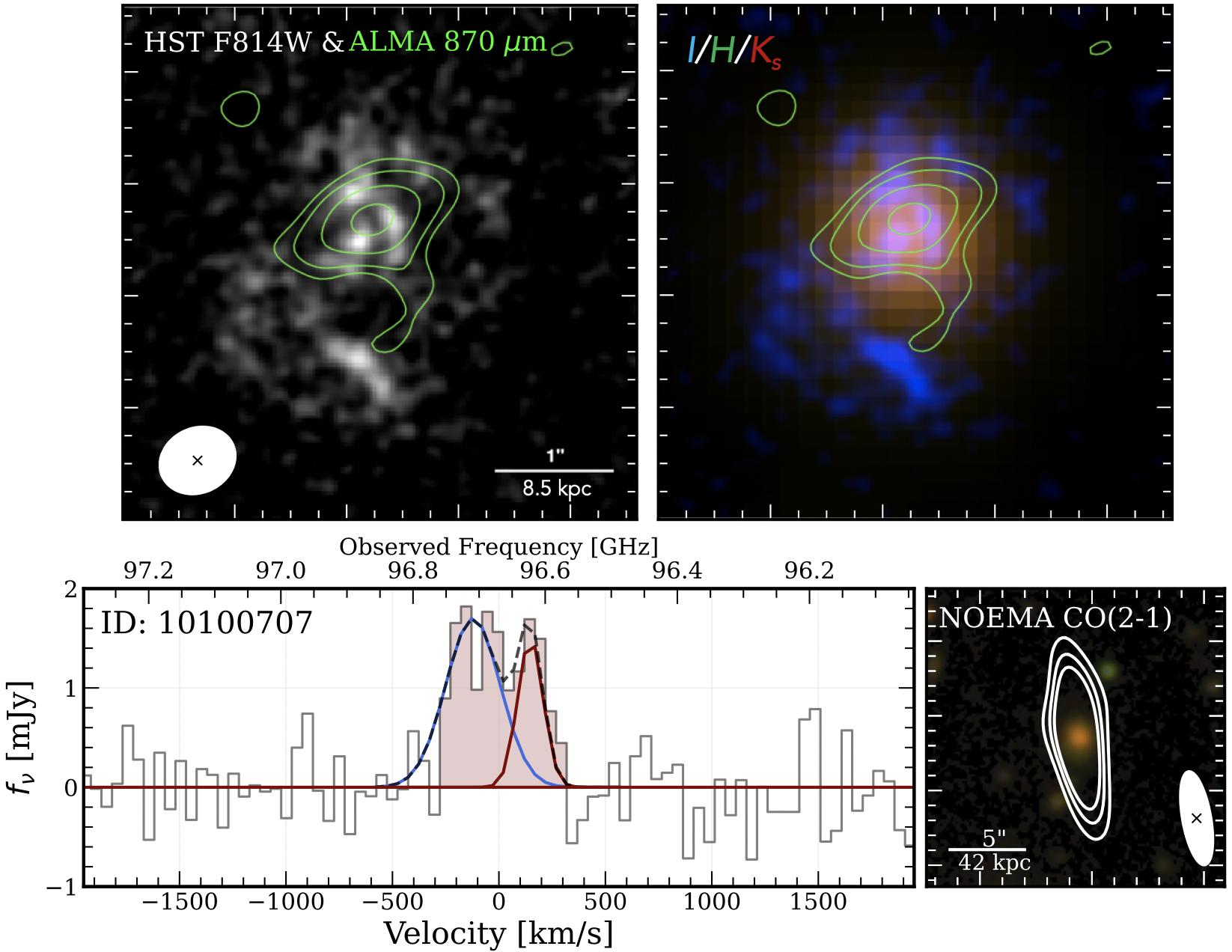}
\caption{'Gas-giant' candidate from our original NOEMA proposal. The galaxy was confirmed at $z_{\rm spec}=1.38$. \textbf{Top Left:} A 5\farcs{0} \textit{HST} F814W cutout of our source with ALMA 870 $\mu$m 2, 3, 5 and 8 $\sigma$ levels shown in green. In the bottom left we show the size of the ALMA beam. \textbf{Top Right:} An RGB composite with blue: \textit{HST}/F814W, green: UVISTA/$H$ and red: UVISTA/$K_{\rm S}$. \textbf{Bottom Left:} Extracted spectra of the CO (2-1) line for our galaxy - ID 10100707. The double Gaussian fit to the spectrum is shown in blue and red, while the sum of the two is represented by a dashed black line.
Shaded maroon region shows the channels used to extract the line flux. \textbf{Bottom Right:} A 20\farcs{0} RGB cutout of our source. We additionally show the 2, 3 and 5 $\sigma$ contours of the CO(2-1) line. The solid white ellipse displays the NOEMA beam size.
}
\label{fig:fig1}
\end{figure*}

The interaction between the interstellar medium (ISM), which consists mostly of gas and dust, and the radiation fields produced by stellar activity, are thought to be the main mechanisms that drive galaxy evolution. Within the contemporary picture of galaxy formation and evolution, star formation rate (SFR), molecular gas masses ($M_{\rm gas}$), dust mass ($M_{\rm dust}$), gas to stellar mass ratio ($f_{\rm gas}\equiv M_{\rm gas}/M_*$) and dust to stellar mass ratio ($f_{\rm dust}\equiv M_{\rm dust}/M_*$) play a critical role in allowing us to interpret the mode and onset of star formation, the assembly of stellar mass, the final quenching of galaxies and their structure and dynamics. In particular, the amount of gas with respect to the ongoing star formation (star formation efficiency, SFE $\equiv$ SFR$/M_{\rm gas}$) and the depletion timescale ($\tau_{\rm depl}\equiv$ 1/SFE) allow us to clearly distinguish the star-formation modes between main-sequence (MS) and starburst (SB) galaxies. It is still unclear, however, whether enhanced SFE, galaxy mergers or increased molecular gas reservoirs are responsible for the manifestation of galaxies in the SB regime.
To this end, the evolution of $f_{\rm dust}$ and $f_{\rm gas}$ across the cosmic time has been extensively studied and constrained both observationally (see \citealt{magdis2012,santini2014,sargent14,genzel15,tacconi18,liu19b,donevski20,magnelli20,kokorev21}), as well as theoretically (e.g. \citealt{tan14,lagos15,lacey16}). 
The important takeaway point of these studies is that both $f_{\rm dust}$ and $f_{\rm gas}$ increase slowly from $z=0$ to their peak at $z\sim2-3$, mirroring the evolution of star formation rate density (SFRD; \citealt{madau14}). 
\par
The other key parameter in studying the evolution of galaxies is the metallicity ($Z$). Metals are introduced into the ISM by either stellar winds and/or via the injection by supernovae \citep{dwek1980,kobayashi20}. 
In this context the gas-to-dust mass ratio ($\delta_{\rm GDR}$) connects the amount of metals locked in the gas phase, with the metals present as dust, therefore acting as a powerful tool to elucidate the evolutionary stage of a galaxy. Observations of both atomic (HI) and molecular hydrogen (H$_2$) have revealed that $\delta_{GDR}$ decreases as a function of metallicity, at least for the local galaxies, (e.g. see \citealt{remy-ruyer14}), however the exact evolutionary scenarios are still uncertain.
\par
Critical to understanding both the metallicity and modes of star formation is how we derive the \mgas. The most commonly used techniques involve utilising the carbon monoxide (CO) or neutral carbon [CI] line luminosities \citep{papadopoulos04,bolatto13,carilli13,valentino18} or the dust continuum (e.g. \citealt{magdis2012,scov2016}). Related to the dust method, in particular, the steadily growing number of galaxy populations with well-sampled IR properties has produced a series of unexpected results. One such discovery is an unusually large \fdust\, (e.g. \citealt{tan14,kokorev21}), in excess of what would be possible to produce involving return fractions from evolved stellar populations \citep{bethermin15,michalowski15,dayal22}. Far IR (FIR) galaxy spectral energy distributions (SEDs) are generally modelled under the assumption of optically thin FIR emission. However, in cases where FIR dust emission is truly optically thick (e.g. the case of Arp220; \citealt{scov17arp220}), using an approximation for optically thin dust results in colder dust temperatures ($T_{\rm dust}$), and therefore overestimated \md and \mgas\, \citep{hodge16,simpson17,scov17arp220,cortzen20,jin22}. Alternatively, some use a constant mass-weighted temperature instead ($T_{\rm dust} \sim 25$ K), as suggested by \citet{scov14}. Recent results by \citet{harrington21}, however, find little difference between mass and luminosity weighted temperatures at least at 1 mm, suggesting that 25 K might be insufficient. Given there is an adequate mid-to-far-IR coverage, it is possible to model galaxies with a more complex, optically thick assumption. However discriminating between the optically thick and thin solutions, when the data are limited, is often not possible due to degeneracies involved in the fitting. As such, to robustly determine the \mgas, dust continuum alone is often insufficient, and independent proxies are thus required to break these degeneracies. 

\begin{figure*}
\centering
\includegraphics[width=0.98\textwidth]{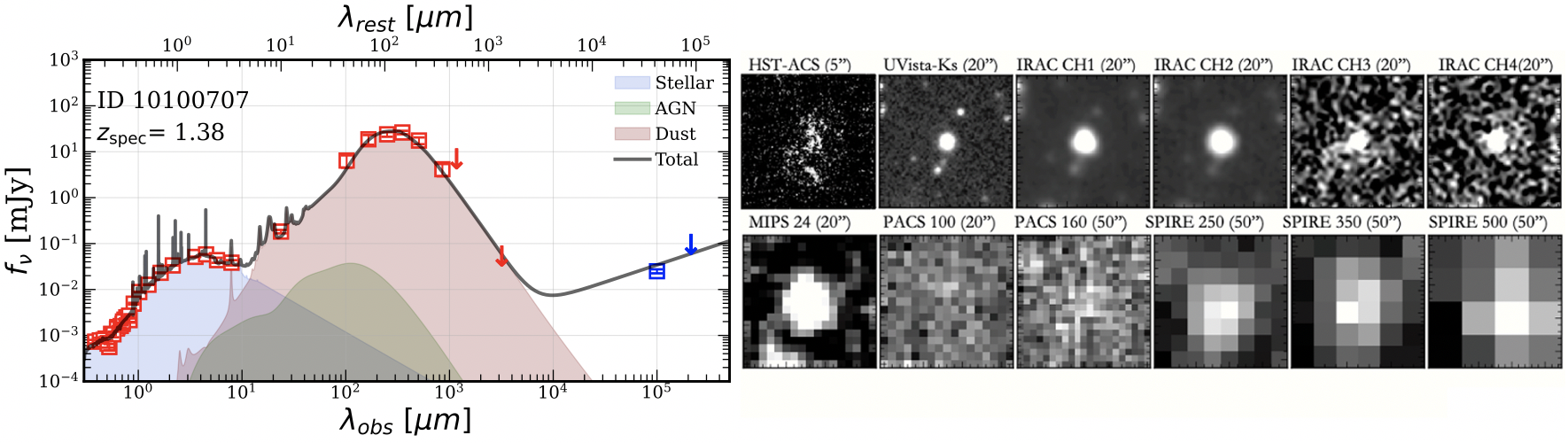}
\caption{\textbf{Left:} Best fit multi-component template \textsc{Stardust} spectrum (grey line and shaded areas), the observed photometry (red squares), including upper limits (red arrows) are also displayed. Radio measurements from VLA at 1.4 GHz \citep{schinnerer2010} and 3 GHz \citep{smolcic2017} are overlaid in blue. The SED is separated into three components - stellar emission (blue), AGN emission (green), dust (red).  \textbf{Right:} Optical-NIR-FIR cutouts of our galaxy, ranging from 5\farcs{0} in the optical-MIR range to 50\farcs{0} in FIR.}
\label{fig:sed_cut}
\end{figure*}

In our homogeneous analysis of the "Super-deblended" catalog \citep{kokorev21} in the COSMOS field \citep{laigle2016,sdc2}, we have recovered scaling relations for FIR properties from thousands of galaxies. Among those we have identified some outliers which appear to have significantly elevated \fdust\, and \fgas, compared to the typical scaling relations at $z<3$ and log$_{10}(M_*/M_\odot) \sim 10.7$. Alternative gas mass tracers are thus necessary to confirm the extreme nature of these objects.  In this paper we report observations of the 
CO (2-1) transition line with Northern Extended Millimeter
Array (NOEMA), for one such "gas-giant" candidate, presented in \citet{kokorev21}.
\par
Throughout this work we assume a flat $\Lambda$CDM cosmology with $\Omega_{\mathrm{m},0}=0.3$, $\Omega_{\mathrm{\Lambda},0}=0.7$ and H$_0=70$ km s$^{-1}$ Mpc$^{-1}$, and a \citet{chabrier} initial mass function (IMF) within $0.1-100$ $M_{\odot}$.

\section{Selection, Data and Reduction} \label{sec:datred}
\subsection{'Gas Giants'}
The sample of 'gas-giants' was initially identified as a number of extreme outliers from the average \fdust\ and \fgas\ evolutionary trends. These typically have log(\fgas) $>$  0.5, i.e. their gas mass reservoir takes $\sim75\%$ of total baryonic matter budget. Individual best - fit \textsc{Stardust} \citep{kokorev21} SEDs of these objects have been examined along with the cut-out images in order to rule out either bad or insufficient optical photometry - resulting in an incorrect $M_*$ estimate, poor coverage of the FIR peak or blending issues that could result in erroneously large \mgas\ estimates. These sources are in stark contrast when compared to the expected evolutionary tracks, with the 'gas-giant' sample being elevated by at least a factor of 3$\times$ in both \fdust\, and \fgas\, compared to median empirical trends, and appearing as $\sim 5\sigma$ outliers to the parent sample. When contrasted with simulations the results are even more puzzling, with \fdust\, deviation of a factor 5$\times$ and a staggering factor of 10$\times$ in \fgas\,, when parameterized to the same SFR/SFR$_{\rm MS}$, $M_*$ and $z$. The possible explanations for the very high \md\ and subsequently \mgas\ estimates, could be optically thick FIR emission, poorly deblended FIR photometry, or simply an erroneous gas-to-dust ratio ($\delta_{\rm GDR}$). All three of these hypotheses have been thoroughly tested, via simulations, and alternative SED fitting techniques, but failed to demote a significant fraction of the "gas giants". We refer the reader to \citealt{kokorev21}, for a more detailed description of the sample.
\par
In an attempt to explain the unusually elevated \fdust\, and \fgas\, we have selected a robust subsample of 'gas-giants' candidates to follow up. These fulfil the following criteria: 1) Secure FIR coverage with $\geq3$ continuum detections above $3\sigma$ level, and at least a single rest-frame $\lambda > 150$ $\mu$m detection for a reasonable $M_{\rm dust}$ and $M_{\rm gas}$ estimates (e.g see \citealt{berta16,kokorev21}), 2) The galaxy needs to be isolated, to facilitate a secure $M_*$ estimate and minimize the possibility of blending in the FIR. From this subsample we selected a single object, to act as a pilot study into this potential population. Using the $M_*$ and SFR together with the $\delta_{\rm GDR}$ - metallicity relation from \citet{magdis2012}, and the \citet{PP04} scaling, we derive a $\delta_{\rm GDR}=95$, computing $M_{\rm gas}=\delta_{\rm GDR} \times M_{\rm dust}=10^{11.49\pm0.20} M_{\odot}$. Our galaxy exhibits what can be considered `typical' values as MS galaxies at log($M_*/M_\odot)\sim 10.9$, and also does not appear to be a strong starburst with SFR/SFR$_{\rm MS} \sim$ 2.5, thus making it even more unique and puzzling. 

\subsection{NOEMA Observations}
To confirm or rule out the high $M_{\rm gas}$, we have used IRAM NOEMA to conduct observations of the CO (2-1) line transition a COSMOS galaxy ID: 10100707 \citep{sdc2}, with a photometric redshift of $z_{\rm phot}=1.35$ from the COSMOS2015 catalog \citep{laigle2016}. The NOEMA observations took place in April 2022, for a total of 8.6 h, using the 12D configuration (Program W21CO, PIs: V. Kokorev, C.G\'omez-Guijarro). In our galaxy the CO(J=$2-1$) line ($\nu_{\rm rest}= 230.5$ GHz) is redshifted to $\nu_{\rm obs}=98$ GHz at $z_{\rm phot}=1.35$. The pointing was centered at the coordinates included in the parent "Super-deblended" catalog \citep{sdc2}. While the D-configuration has a generally lower angular resolution ($\sim 3-6$"), our original proposal was focused on the line detection and thus was the most optimal balance between S/N and requested time.  The source was observed with three tracks reaching rms
sensitivities of 0.13 mJy per 500 km s$^{-1}$.
\par
The separate tracks were calibrated and recombined into a single $uv$ table, using the \textsc{GILDAS} software package \textsc{CLIC}. No continuum was detected however we use the information from the separate tracks to derive the upper limit on the underlying 98 GHz (3 mm) continuum, which will further assist in deriving robust FIR properties from SED fitting. To produce the final 1-D spectra for the CO(2-1) line, we have used the \textsc{Crab.Toolkit.PdBI} \footnote{\url{https://github.com/1054/Crab.Toolkit.PdBI}}, which is a wrapper around the \textsc{MAPPING} module in \textsc{GILDAS}. Finally, we fit a point source model in the $uv$ space at the position of peak flux in order to measure the flux density of the line. The line profile presented in \autoref{fig:fig1} appears to be double peaked. Due to that, in addition to the single Gaussian peak fit, we also model the line profile with two normal distributions.

\subsection{ALMA Data}

The galaxy has been observed with ALMA band 7 at 341 GHz (870 $\mu$m) as a part of the program 2011.0.00097.S (PI: N. Scoville). Following the method in \cite{jin19,jin22}, we reduced and calibrated the raw data using CASA pipeline. The calibrated data were exported into uvfits format to generate $uv$ tables of the IRAM GILDAS, after which we perform further analysis in the $uv$ space (visibility).
The continuum map of the galaxy was imaged by combining $uv$ visibilities of all spectral windows using \texttt{uv\_average} and \texttt{clean} procedures in GILDAS/Mapping.
The clean continuum image has an rms of 114.3~$\mu$Jy and a peak flux of 1.096~mJy. The synthesised beam is $0.65''\times0.54''$ with a position angle of $-65^{\circ}$. 

As the contours show in \autoref{fig:fig1}, the ALMA 870 $\mu$m image is clearly resolved, as evidenced by the larger size than the synthesized beam. Therefore, we further fit the dust continuum with an elliptical Gaussian model in $uv$ space using GILDAS \texttt{uv\_fit}, which gives a total flux of $1.81\pm0.26$ mJy and a size of $(0.65\pm0.11)''\times(0.32\pm0.10)''$ (PA$=-90^{\circ}$).  The total flux is higher than the peak flux density, indicating that the source is spatially resolved.

\begin{table}
\def\arraystretch{1.2}
\caption{Description of the Source}
\centering
\label{tab:general-fir}
\begin{tabular}{cc}
\hline\hline
ID & 10100707 \\
\hline
R.A. &  150.554\\
Dec &  2.422 \\
$z_{\rm phot}$  & $1.35\pm0.03 $\\ 
$z_{\rm CO}$ &  1.3844\\ 
$A_{\rm V}$ & $1.81\pm0.05$ \\
log($L_{\rm IR,dust}/L_{\odot}$) & $12.09\pm0.02$\\
SFR [M$_{\odot}$ yr$^{-1}$]  &  $123\pm6$\\
log($M_\ast/M_{\odot}$) &  $10.91\pm0.06$ \\
log($M_{\rm dust}/M_{\odot}$) &  $9.25\pm0.21$ \\
SFR/SFR$_{\rm MS}$ $^{1}$ &  $1.74\pm0.15$ \\
$T_{\rm dust,thin}$ [K] &  $26.2\pm2.2$ \\
$T_{\rm dust,thick}$ [K] &   $46.5\pm3.1$ \\
log($M_{\rm gas}/M_{\odot})_{\rm Dust}$ &  $11.20\pm0.20$ \\
12+log(O/H) $^2$ & $8.67\pm0.15$ \\
\hline \hline
$S_{\rm CO(2-1)}$ [Jy km$^{-1}$]   &  0.78$\pm$0.05  \\
FWHM$_{\rm CO(2-1)}$ [km s$^{-1}$]    & $442\pm50$  \\
$L$'$_{\rm CO(2-1)}$ [K km s$^{-1}$ pc$^2$] & $2.05\times10^{10}$\\
log($M_{\rm gas}/M_{\odot})_{\rm CO}$ & $10.84\pm$0.03 \\
$r_{\rm eff, K_{\rm S}}$ & $0\farcs{6}\pm0\farcs{1}$ \\
$r_{\rm eff, {\rm ALMA}}$ & $0\farcs{23\pm0\farcs{06}}$ \\
log($M_{\rm dyn}/M_{\odot}) {K_{\rm S}}$ & $11.32\pm0.10$ \\
log($M_{\rm dyn}/M_{\odot}) {\rm ALMA}$ & $10.87\pm0.10$ \\
$\delta_{\rm GDR}$ &  $39\pm10$ \\
\hline \hline
\end{tabular}
\begin{tablenotes}
\item[$^1$] \footnotesize{Assuming the MS relation from \citet{schreiber15}. \hfill
\item[$^2$] \footnotesize{Metallicity is expressed in the \citealt{PP04} scale.} \hfill}
\end{tablenotes}
\end{table}

\section{Results} \label{sec:results}
 We present the resultant 1-D spectra in the bottom panel of \autoref{fig:fig1}, along with the RGB image which was created by combining the $HST$/ACS F814W, UltraVista DR2 (UVISTA; \citealt{uvista}) $H$ and $K_{\rm s}$ filters. For our object we compute the integrated CO (2-1) line flux by taking the product of the average flux density in the appropriate channels, maximising the SNR and the velocity in these channels (e.g. see \citealt{daddi15}). Generally line flux estimates are performed by modelling the emission with a Gaussian profile, we however compared our non-parametric approach to the Gaussian fit and find results to be consistent.

\begin{figure*}
\centering
\includegraphics[width=0.98\textwidth]{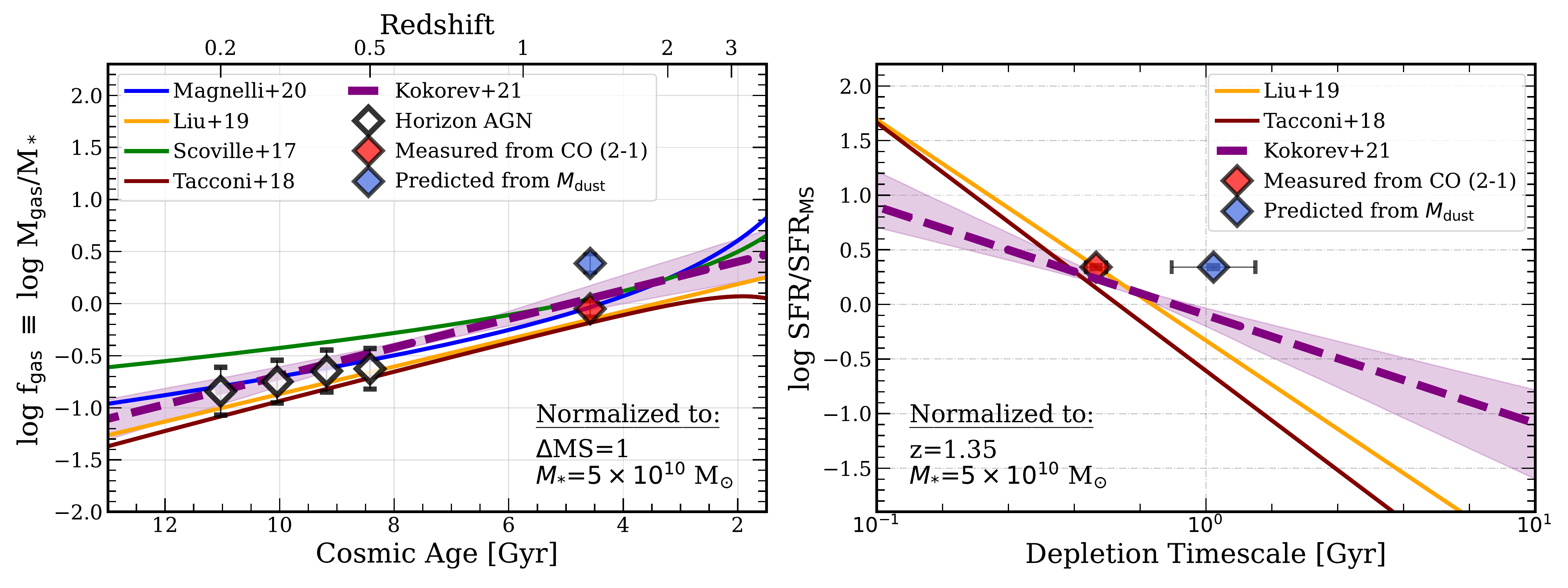}
\caption{ \textbf{Left:} Evolution of $f_{\rm gas}$ as a function of $z$/cosmic age. Values of $M_{\rm gas}$ derived from $M_{\rm dust}$, and \coto\, are shown as red and blue diamonds respectively. Downward arrows denote upper limits. Solid colored lines correspond to empirical evolutionary trends derived in \citet{scov17},\citet{tacconi18}, \citet{liu19b}, \citet{magnelli20}. The dashed purple line shows the fit to the COSMOS Super Deblended catalogue data, as described in \citet{kokorev21}, with shaded purple region denoting the 16$^{\rm th}$ and 84$^{\rm th}$ percentile confidence intervals. Both the data and the derived relations have been re-scaled to $\Delta$MS=1 and $M_*=5\times10^{10}$ M$_\odot$. White diamonds show median positions of the Horizon AGN star-forming galaxies at that redshift \citep{laigle19}, normalised in the same way as our data. 
\textbf{Right:} The change in the distance to the main-sequence as a function of depletion time ($\tau$). Labels and lines are the same as on the first panel.}
\label{fig:fgas_tau}
\end{figure*}

While it quickly becomes uncertain for high ionisation states of CO, the conversion factor between \coto\ and \cooz\ is small and very well constrained, it is also applicable for normal BzK galaxies on the MS, and presumably even more extreme objects which we describe in this work \citep{daddi15}. We therefore convert the \lcot\, to \,\lco\, by assuming a standard CO SLED  with the conversion factor \lcot/\lco = $r_{21}$ = 0.85 \citep{bothwell13}.

\subsection{Revised Gas Mass}
The best-fit SED, along with the cutouts for the galaxy are shown in \autoref{fig:sed_cut}. This galaxy lies on the main sequence of star formation, as parametrised in \citet{schreiber15} with SFR/SFR$_{\rm MS}$<5. Therefore, to convert the final value of \lco\, to \mgas\, we assume $\alpha_{\rm CO}=4.0$ \citep{bolatto13}. In \autoref{tab:general-fir} we present the final results from our \coto\ observations, alongside other optical and FIR properties computed with \textsc{Stardust} in our previous work. 
\par
This 'gas-giant' was initially identified due to its elevated \fdust\, rather than \fgas\,, with the former being robustly computed from the Herschel Space Observatory ($Herschel$) and SCUBA2 detections, which securely cover the Rayleigh-Jeans (RJ) tail ($>250$ $\mu$m, rest-frame). Using the additional constraint on the upper limit derived from the underlying 3 mm continuum, and the secure $z_{\rm spec}$ we re-fit our object to ensure that the previously derived properties are robust. We find that all FIR parameters remain unchanged from the ones presented in \citet{kokorev21}, apart from the $M_{\rm dust}$ which has fallen by $0.3$ dex, from log$(M_{\rm dust}/M_{\odot})={9.54\pm0.10}$ to log$(M_{\rm dust}/M_{\odot})={9.25\pm0.21}$. As such we also update the previous estimate of the sub-mm continuum derived $M_{\rm gas}$ from log$(M_{\rm gas}/M_{\odot})={11.49\pm0.20}$ to log$(M_{\rm gas}/M_{\odot})={11.20\pm0.20}$. The results from the revised SED fitting are, however, still insufficient to bring our galaxy on the \fdust\, and \fgas\ evolutionary tracks.
\par
In a stark contrast to the updated \mgas\, which we compute from re-fitting the FIR data, the CO derived mass is found to be lower by 0.4 dex. We display the updated physical parameters with respect to the \fgas\, vs cosmic age and distance to the main sequence (SFR/SFR$_{\rm MS}$) vs $\tau$ relations of our source in \autoref{fig:fgas_tau}, additionally contrasting it to the updated SED fitting predictions of the \mgas. The new CO-based \mgas\, value has resulted in a decrease of both \fgas\, and $\tau$, thus placing our object within the range of values expected from observational results \citep{scov17,tacconi18,liu19b,magnelli20,kokorev21}, as well as simulations \citep{laigle19}. As a result our new observations now can rule out this object as a "gas-giant" along with possible physical scenarios discussed in \citet{kokorev21}.
\par
The elevated \fdust\, still remains in disagreement with the existing literature trends, and now, in light of our CO(2-1) observations, creates even more tension  with the updated \mgas\, estimate. For example, as displayed in \autoref{fig:gdr}, the re-computed $\delta_{GDR}$ is a factor of $\sim 2.5\times$ lower compared to our previous prediction. We will discuss the potential reasons for this newfound tension in the next section.

\subsection{Morphology}
Compared to the deblended SCUBA2 850 $\mu$m flux ($S_{850}=4.16\pm1.24$ mJy), the total ALMA 870 $\mu$m flux is lower by a factor of 2.3. In order to capture the total flux, we also adopted an aperture of $2\farcs{5}\times2\farcs{0}$ with the same PA as Gaussian model, which results in a maximum flux of $2.05\pm0.31$~mJy, still lower than the SCUBA2 flux by a factor of 2. We argue, however, that the high SCUBA2 flux is not due to blending, because this galaxy is the only source visible in ALMA primary beam ($20\farcs{0}$). The SCUBA2 beam is $15\farcs{0}$, and thus is not contaminated by a dusty neighbor. The SCUBA2 photometry itself is well-measured using the $K_{\rm s}$ prior position, with the uncertainty calibrated using MonteCarlo simulations \citep{sdc2}. The only reasonable explanation of the flux discrepancy is that ALMA has over-resolved the galaxy, losing more than half of the total flux due to the incomplete $uv$ coverage. In detail, the ALMA max recovery scale of this pointing is only $4.7"$, the emission from larger scale would be largely lost. This resolved nature is again confirmed by the extended morphology as present in both NIR and ALMA images. The significant flux loss in ALMA also hints that the dust emission is more extended out to $>1\farcs{2}$ ($\sim 10$ kpc), which is consistent with the sizes reported in \citep{valentino20} for a sample at $\langle z \rangle =1.2$. Given the linear scaling of $M_{\rm dust}$ with flux in the RJ tail, we find that ALMA continuum ends up resolving out more than half of the total $M_{\rm dust}$ reservoir.
\par
In \autoref{fig:fig1} we show the ALMA 870 $\mu$m together the RGB composite consisting of the F814W, UVISTA-$H$ and UVISTA-$K_{\rm s}$ bands. We find that while our galaxy is not fully detected in F814W it appears to display a clumpy morphology. In order to estimate the optical size of the galaxy we therefore fit the UVISTA/$K_{\rm s}$ band image with \texttt{IMFIT} \citep{erwin15}, using a \citet{sersic} profile. In addition to that we verify our result by computing the band $r_{\rm eff}$ from the mass-size relation presented in \citet{vanderwel14}. Both methods return a consistent result of $r_{\rm eff}=0\farcs{6}\pm0\farcs{1}$. We also compute the ALMA size by converting the elliptical axes measured by \textsc{GILDAS} into $r_{\rm eff}=0\farcs{23\pm0\farcs{06}}$. At $z=1.38$ we thus get a final size of $5.1\pm0.8$ kpc in $K_{\rm S}$ band, and $1.9\pm0.4$ kpc in ALMA. However, while the peak of the ALMA emission coincides with the apparent center of the galaxy, we also report the tentative (2$\sigma$ level) and large ($\sim 10$ kpc) dust arm. This dusty structure appears to be connected to the secondary clump seen in the south part of the F814W image. It is likely that the dusty arm is associated with the smaller peak in the CO (2-1) profile, however further high resolution observations would be required to verify this.

\subsection{Dynamical Mass}
Using the fact that the \coto\ line is detected in multiple channels, we also compute the dynamical mass ($M_{\rm dyn}$) for our object. To do that we consider both the UVISTA/$K_{\rm s}$ and ALMA sizes measured in the previous section.
We use the $M(r<r_{\rm eff})$ relation from from \citet{daddi10} and estimate an inclination angle for our galaxy to be $i=(60\pm4)^{\rm o}$ from the axis ratio measured on the ALMA 870 $\mu$m image. This is also consistent with the statistical average of $(57\pm21)^{\rm o}$ \citep{coogan18}.
The total $M_{\rm dyn}$ is then given as the mass contained within the full diameter of the galaxy - $2\times M(r<r_{\rm eff})$. We assume the contribution of dark matter to be $\sim 10\%$, as there is evidence that dark matter fraction at this redshift is negligible within the $r_{\rm eff}$ \citep{daddi10,wuyts16,genzel17}.
\par
We find dynamical masses of log$_{10}$($M_{\rm dyn}/M_{\odot})=11.32\pm0.10$ and log$_{10}$ ($M_{\rm dyn}/M_{\odot})=10.87\pm0.10$, using the optical and ALMA $r_{\rm eff}$~ respectively. Given our previous discussion, it is very likely that the galaxy is over-resolved at 870 $\mu$m, and thus its size and $M_{\rm dyn}$ would be underestimated, which is in line with $M_{\rm dyn, ALMA}$ being lower than the $M_*$. We therefore based our $M_{\rm dyn}$ on the $K_{\rm s}$ size of the galaxy. While it is rare for the sub-mm size to extend as far as the star-forming stellar disk, probed by the $K_{\rm s}$ band (e.g. see \citealt{puglisi21,gomez-guijarro22}), this $M_{\rm dyn}$ estimate can serve as a robust upper limit. The $M_{\rm dyn}$ are listed in \autoref{tab:general-fir} without the dark matter correction.

\section{Discussion}
\subsection{Stellar Mass Underestimate}
One of the potential reasons behind the "gas-giant" population discussed in \citet{kokorev21} is an underestimate in $M_*$. In return leading to erroneously larger dust/gas-to-stellar mass ratios. The optical-NIR data that was used to derive the $M_{*}$ relies mostly on the ground-based observations. This can result in potential flux losses, as a result of seeing limited point spread function (PSF). The \citet{laigle2016} catalog used in our work utilises apertures to measure flux density, and might not adequately model the flux losses due to the wings of the PSF  (e.g. see \citealt{weaver22}). \par
Despite that, when using the CO derived $M_{\rm gas}$ to compute $M_{\rm gas}/M_*$, our galaxy is in line with the literature predictions for a MS galaxy at that redshift \citep{scov17,tacconi18,liu19b,magnelli20,kokorev21}, as well as an extrapolation of Horizon AGN simulations results to higher-$z$ \citep{laigle19}. Therefore we have little reason to believe that an $M_*$ underestimate is responsible for the elevated $M_{\rm dust}/M_*$.

\begin{figure*}
\centering
\includegraphics[width=1\textwidth]{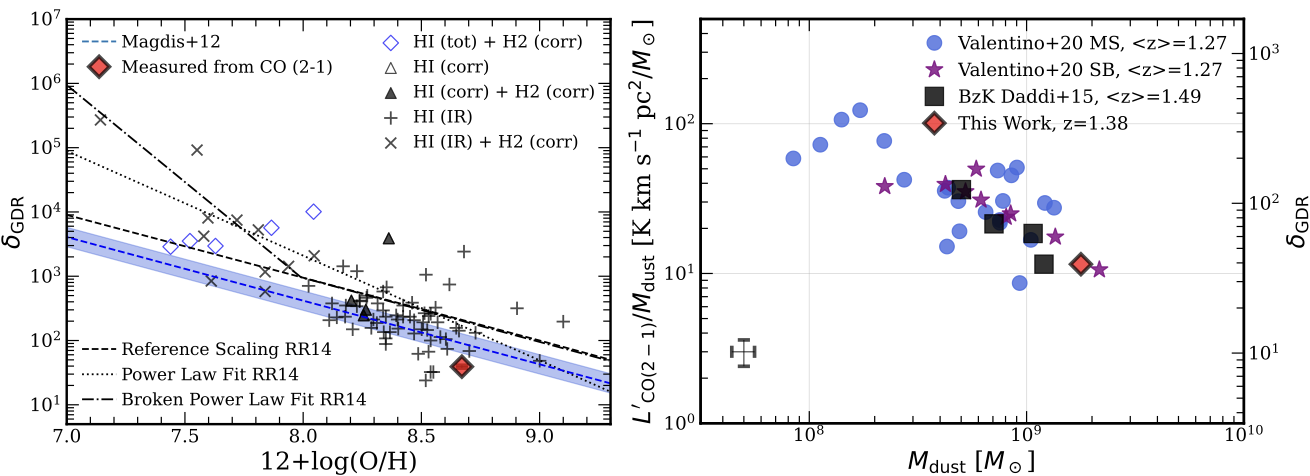}
\caption{\textbf{Left:} Gas-to-dust ratio as a function of metallicity for the galactic CO to molecular gas conversion from \citet{remy-ruyer14} (RR14). The dashed line represents the reference scaling of the $\delta_{\rm GDR}$ with metallicity and the dotted and dash-dotted lines represent the best power-law and best broken power-law fits to the data presented in RR14. The dashed blue line with the shaded envelope show the $\delta_{\rm GDR}$ - metallicity relation from \citet{magdis2012} which we used to derive the original $M_{\rm gas}$. The $\delta_{\rm GDR}$ measurement from CO and dust continuum is shown as a red diamond. \textbf{Right:} A CO(2-1) line luminosity to $M_{\rm dust}$ ratio and $\delta_{\rm GDR}$ as a function of $M_{\rm dust}$. For the $\delta_{\rm GDR}$ we assume $\alpha_{\rm CO}=4.0$ for all galaxies. We also show the sample of MS (blue) and SB (purple) galaxies from \citet{valentino20}. The BzK galaxies from \citet{daddi15} are shown as black squares. Our galaxy is shown as a red diamond. A typical uncertainty on the data is shown in the bottom left corner.}
\label{fig:gdr}
\end{figure*}

\subsection{Optically Thick Dust}
Another mechanism responsible for the elevated $M_{\rm dust}$/$M_*$ found in our galaxy could be the presence of optically thick FIR emission. Indeed in \citet{kokorev21} we report that ID: 10100707 can be fit with a generalized form of the modified blackbody (MBB) function (e.g. see \citealt{casey12}). In that case we find the effective
wavelength ($\lambda_{\rm eff}$) at which the optical depth $\tau$ becomes unity, to be $\lambda_{\rm eff}\sim600$ $\mu$m, deep in the RJ tail. In this case the intrinsic SED is shifted blueward, reducing the derived $M_{\rm dust}$ by a factor of a few. The unusually high dust-to-stellar mass ratio and $\delta_{\rm GDR}$ when using the CO derived $M_{\rm gas}$ can therefore be a result of the optical depth effects. As we report in \citet{kokorev21}, however, adopting an optically thick dust emission model reduces the $M_{\rm dust}$ by 0.25 dex, compared to the 0.4 dex correction required, which is insufficient to bring our galaxy back up to the expected dust-to-stellar mass ratio relation which we show in \autoref{fig:fgas_tau}.

\subsection{Dust and Metallicity}
In \autoref{fig:gdr} we examine the location of our source on the $\delta_{\rm GDR}$ vs $Z$ relation. We additionally overplot the $\delta_{\rm GDR}$-$Z$ relation from  \citet{magdis2012}, which was originally used to compute the original $M_{\rm gas}$ as well as the relations from \citet{remy-ruyer14}. Our findings indicate that ID: 10100707 displays a $\delta_{\rm GDR}$ that is 2.5 times lower than what is expected from the scaling relations for star-forming galaxies, given the metallicity we have estimated ($\sim Z_\odot$). For our galaxy to follow the scaling relations with its $\delta_{\rm GDR}$, it would require a super-solar metallicity of $>3\times Z_\odot$. Although it is possible that our object is simply very metal-rich, confirmation would require more direct metallicity indicators.
\par
Conversely, an increasing number of theoretical studies highlight the significance of grain growth in dust production within the ISM, particularly for metal-rich galaxies (e.g. \citealt{asano13,remy-ruyer14,hirashita15,zhukovska16,devis19}). Depending on the star-formation history within the galaxy, grain growth can become the dominant mechanism of dust production once a certain metallicity threshold is reached. As a result, metals are depleted from the ISM as the $M_{\rm dust}$ increases significantly (also see \citealt{donevski23}). Therefore, it is possible that our galaxy indeed has solar metallicity, and the observed $\delta_{\rm GDR}$
may be a result of rapid metal-to-dust conversion via grain growth.
\subsection{Dust to Gas Conversion}
From our discussion and analysis we predict that the systematic shifts induced on $M_*$ and $M_{\rm dust}$ are not applicable or insufficient to explain the derived $\delta_{\rm GDR}$. Generally, the metallicity is not the only driver of the observed scatter in the $\delta_{\rm GDR}$ values, and other phenomena in galaxies can lead to a large variation of this parameter. These would include the morphological type, stellar mass and SFR, however each of them having an effect on the observed $\delta_{\rm GDR}$ \citep{remy-ruyer14}.
\par
It is understood that when deriving the $M_{\rm gas}$ via the $M_{\rm dust}$ method, what is actually being computed is the total mass of hydrogen HI+H$_2$ (e.g. see \citealt{magdis2012}). It is therefore possible that our galaxy has elevated HI reservoirs, not traceable by CO, that could cause the observed discrepancy. By combining our dust and CO based $M_{\rm gas}$ measurements we compute the HI to stellar mass ratio $M_{\rm HI}/M_*\sim 1$, which is roughly 2 times larger than expected at that redshift \citep{heintz22}. Using the 21-cm line emission, however, \citet{chowdhury20,chowdhury21} find that 
at $z\sim1$ some galaxies can indeed reach $M_{\rm HI}/M_*$ similar to, and in excess, of our findings. A large HI reservoir is therefore a plausible explanation for the elevated $\delta_{\rm GDR}$.
\par
The observed disparity between the $\delta_{\rm GDR}$ - $Z$ relation for our object could also be explained by the presence of the CO - dark molecular gas. In this case the H$_2$ reservoirs which can not be traced by CO (e.g. \citealt{papadopoulos02,rollig06,wolfire10,glover12,madden20}) will require another method of tracing optically thick gas not normally probed by CO. In this case, hydrodynamical simulations (e.g. see \citealt{smith14,offner14,franeck18,seifried20}) have shown that neutral [CI] and ionized [CII] carbon, have the ability to trace molecular hydrogen in optically thick and warm environments. Comparably, the dust in our object is both optically thick out to $\sim 600$ $\mu$m, and is warmer than in a typical MS galaxy at that redshift (e.g. \citealt{schreiber15,schreiber18}). It is therefore possible that our CO (2-1) study fails to quantify the full H$_2$ reservoir, and further [CI] observations would be necessary to correctly compute the $M_{\rm gas}$. Moreover, while optically thick HI has been suggested as a potential source of CO-dark gas, there is currently no convincing evidence to support its significant contribution to the dark neutral gas \citep{murray18}. \par
Based on the CO $M_{\rm gas}$ estimate, the galaxy can no longer be classified as a "gas-giant", yet the $M_{\rm dust}/M_*$ ratio remains high. We find that the $M_{\rm gas}/M_*$ ratio is marginally consistent with expected values, while the dust reservoir and $\delta_{\rm GDR}$ are abnormal compared to galaxies of similar mass and redshift. Therefore, it would be more accurate to refer to ID: 10100707 as a "dust-giant".
\par
Finally, in the right panel of \autoref{fig:gdr} we examine the position of our galaxy on the \lco$/M_{\rm dust}$ and $\delta_{\rm GDR}$ vs $M_{\rm dust}$ diagram. We find that our galaxy is not an outlier compared to either the MS and SB sources from \citet{valentino20}, as well as the BzK population discussed in \citet{daddi15}. However, we find that our "dust-giant" displays \lcot$/M_{\rm dust}$ ratio consistent with those of SB objects, despite it lying on the main-sequence. In fact the object most closely associated with ID: 10100707 is an extreme starburst (SFR/SFR$_{\rm MS}\sim 9$) of similar $M_{\rm dust}$ and $M_*$ presented in \citet{puglisi21}.

\subsection{Dynamical Mass Budget}
Using the $M_{\rm dyn}$ computed in the previous section we would like to understand whether the derived sub-mm continuum and CO $M_{\rm gas}$ are compatible with the total mass budget. Assuming a dark matter fraction of 10 \% and using the $K_{\rm S}$ galaxy size we obtain the log$_{10}(M_{\rm bar}/M_\odot)=11.25\pm0.10$. While the $K_{\rm S}$ galaxy size is most likely larger than the CO (2-1) emitting region (e.g. see \citep{puglisi21,gomez-guijarro22}) we believe it represents an adequate upper limit on the total baryonic mass of our galaxy, as opposed to the over-resolved ALMA 870 $\mu$m emission. Subtracting the $M_*$ and $M_{\rm dust}$ leaves us with the allowed  log$_{10}(M_{\rm gas}/M_\odot)\leq10.97\pm0.15$, which lies within 1$\sigma$ from our sub-mm and CO $M_{\rm gas}$ estimates. In this case, taking into account the measured \lcot\, this would require an $\alpha_{\rm CO}=4.3\pm0.5$, a value typical of MS galaxies, which aligns with our SFR/SFR$_{\rm MS}$ characterization. Due to the large uncertainty on the $M_{\rm dyn}$ it is however not possible to ascertain which one of $M_{\rm gas}$ estimates is most likely to be correct.

\begin{figure}
\centering
\includegraphics[width=.45\textwidth]{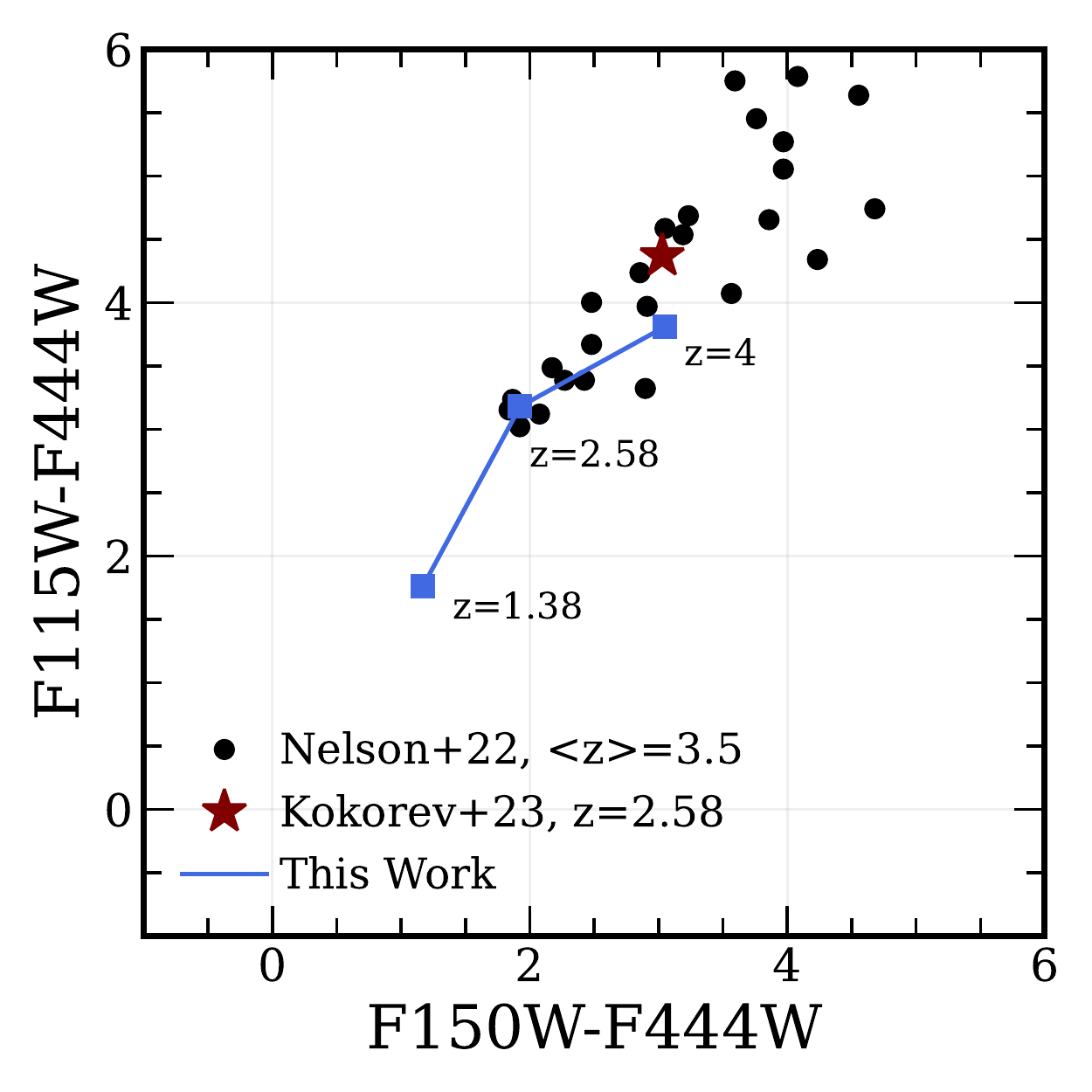}
\caption{$JWST$ color-color diagram for selecting $HST$-dark objects presented in \citet{nelson22}. Black points show the positions of the \citeauthor{nelson22} sources. The maroon star denotes the highly dust obscured $A_{\rm V}\sim5$ galaxy from \citet{kokorev23}. Blue squares denote the location of our object, as well as how it would appear at $z=2.58$ and $z=4$.} 
\label{fig:hst_dark}
\end{figure}

\subsection{Optically-Dark Dusty Contaminant}
Finally, a presence of the optically dark companion, contaminating the FIR-mm emission, could potentially explain the elevated dust content. The existence of these dust obscured, optically faint, star-forming galaxies has been demonstrated by a large number of detections with \textit{Spitzer Space Telescope} (\textit{Spitzer}) and Herschel Space Observatory (\textit{Herschel}; see \citealt{huang11,caputi12,alcpamp19}), as well as in the sub-mm regime \citep{talia21,wang21} specifically by Atacama Large Millimeter/submillimeter Array (ALMA) \citep{simpson14,franco18,yamaguchi19,wang19,williams19,umehata20,caputi21,fudamoto21,manning21,gomez-guijarro22,kokorev22,xiao23}. \par
A potentially similar case was presented in \citet{kokorev23}, where a spatially resolved analysis with James Webb Space Telescope ($JWST$) NIRCam reveals a highly dust obscured disk ($A_{\rm V}\sim5$) at $z=2.58$. Previously only seen as a sub-mm galaxy (SMG) in \citet{laporte17} and \citet{munoz18}, this \hst-dark source has only been possible to detect with deep $JWST$ observations in the NIR. If such an object is indeed boosting the sub-mm flux density,  it would have to be a
FIR/ALMA only source. Further deep observations with $JWST$ would be necessary, in order to confirm if the extreme $\delta_{\rm GDR}$ seen in our galaxy is in fact a result of such a dusty contaminant.

\subsection{Low-$z$ Analog of \textit{HST}-dark Galaxies}
We find that this galaxy displays similar amount of dust obscuration ($A_{\rm V}\sim1.81$), is on the main-sequence and has a comparable $M_*$ and $M_{\rm dust}$ values to the higher-$z$ optically faint sources recently presented in the literature (e.g. \citealt{franco18,wang19,nelson22,barrufet23,gomez-guijarro23,kokorev23}). 
To investigate this possibility, we apply the $JWST$ color-color classification presented in \citet{nelson22}, which uses the F115W-F444W and F150W-F444W colors to identify candidate objects. We integrate our best-fit \texttt{Stardust} template through the $JWST$ filters and obtained the synthetic flux densities. In \autoref{fig:hst_dark} we display the location of our galaxy, along with the \citeauthor{nelson22} objects and the $HST$-dark galaxy reported in \citet{kokorev23}, on the color-color diagram. Furthermore, we trace the position of ID: 10100707 on the diagram if it were located at the same redshift as the \citeauthor{kokorev23} object or at $z=4$. Our analysis suggests that while our object would not qualify as an $HST$-dark galaxy at $z=1.38$, it would meet the criteria for such classification at $z>2$.
\par
Following a similar analysis presented in \citet{kokorev23} we would like to test how likely is it to detect analogs of this object at high-$z$.
We use the detection limits of major $JWST$ surveys, including UNCOVER (PIs: I. Labbe, R. Bezanson; \citealt{bezanson22}), CEERS (PI: S. Finkelstein; \citealt{bagley22}) as well as COSMOS-Web (PIs: J. Kartaltepe, C. Casey; \citealt{casey22}). We also include the $5\sigma$ limit of the deepest public NIRCam observations taken at the moment of writing, from NGDEEP (PI: S. Finkelstein; \citealt{bagley23}). We find that at $z\sim4$ a similarly dusty galaxy would become an F200W dropout with NGDEEP - like depths ($\sim 31$ AB mag), and is \textit{JWST}-dark in all other major surveys. Dusty 2 $\mu$m dropouts with elevated $M_{\rm dust}$ could be misidentified as $z>15$ galaxies, as has been discussed in \citet{naidu22,zavala22,donnan23} and most recently in \citet{arrabalharo23}. We thus highlight that the identification of dust-obscured main-sequence galaxies at and beyond the epoch of reionization remains a challenging task for \textit{JWST}, and these will most likely end up being missed. 

\section{Conclusions}
In this paper we describe NOEMA CO (2-1) line observations and analysis in a galaxy originally selected for its elevated $M_{\rm gas}/M_*$. The object constitutes a pilot study of the "gas-giant" population presented in \citet{kokorev21}. With our new observations we have been able to spectroscopically confirm the redshift of our galaxy, as well as use the line luminosity to derive the $M_{\rm gas}$. Additionally we updated the initial estimate of $M_{\rm dust}$ and dust-based $M_{\rm gas}$ with the addition of the 3 mm continuum upper limit from our NOEMA observations. With the re-computed CO based $M_{\rm gas}$, however, we found that our initial "gas-giant" selection no longer applies. The updated mass estimates, differ by -0.4 dex from the ones computed from $M_{\rm dust}$, and position our galaxy along the expected evolutionary trends of $M_{\rm gas}/M_*$. At the same time we found that the $M_{\rm dust}$ remains elevated, both with respect to the $M_*$ and now $M_{\rm gas}$, leading to unusually low $\delta_{\rm GDR}$ judging from the fundamental metallicity relations for a MS galaxy at this redshift and $M_*$. Given the above it is more appropriate to refer to this galaxy as a "dust-giant" instead. 

Using the additional ALMA data at 870 $\mu$m we find that the dust in our galaxy is extended, which is also supported by the fact that ALMA recovers a lower flux compared to the SCUBA2 data. Coupled with the double peaked CO (2-1) line this hints that gas in our galaxy is present both in the central region, as well as the spiral arms. In addition we note the presence of the blue clumps visible in the F814W image (rest-frame UV).
\par
We find that while modelling our galaxy with optically thick dust emission reduces the dust mass, it is still not sufficient to bring it in line with the scaling relations. Instead, the discrepancy in the physical picture could potentially be explained by enhanced HI reservoirs, CO-dark H$_2$ gas, an unusually high metallicity, or the presence of an optically dark, dusty contaminant. What is clear is that it is important to be cautious when drawing conclusions about the physical picture based solely on $M_{\rm gas}$ derived from dust continuum measurements, given the significant difference between expected and measured values. Problems with the continuum derived $M_{\rm gas}$ are becoming very apparent due to the extreme optically thick dust, as found in \citet{jin22}.
\par
Our object also displays similar amount of dust obscuration compared to the so-called $HST$-dark galaxies which exist at higher-$z$. Using the $JWST$ based color-color selection for such galaxy types we have found that if our galaxy would exist at $z>2$, it would be meet the criteria of such a color selection.  In our work we aim to highlight that the identification of dust-obscured MS galaxies at and beyond the epoch of reionization remains a challenging task for $JWST$, and a the most dusty galaxies will end up being missed.
\par
While the source analyzed and discussed in our work will not be imaged as a part of COSMOS-Web $JWST$ survey \citep{casey22}, we expect that a few tens of galaxies that belong to the "gas-giants" population will still be covered. Future observations with $JWST$/NIRCam and MIRI instruments as a part of COSMOS-Web, as well as deep multi-band sub-mm data would therefore be imperative to understand the true nature of objects with elevated dust-to-stellar mass ratios.

\begin{acknowledgements}
We thank Pratika Dayal for her helpful suggestions which helped improve this manuscript. We would like to thank Orsolya Feher for her help with NOEMA data reduction process. Based on observations carried out under project number W21CO with the IRAM NOEMA Interferometer [30-meter telescope]. IRAM is supported by INSU/CNRS (France), MPG (Germany) and IGN (Spain). 
SJ is supported by the European Union's Horizon Europe research and innovation program under the Marie Sk\l{}odowska-Curie grant agreement No. 101060888.
The Cosmic Dawn Center is funded by the Danish National Research Foundation under grant No. 140. GEM acknowledges the Villum Fonden research grant 13160 “Gas to stars, stars to dust: tracing star formation across cosmic time,” grant 37440, “The Hidden Cosmos,” and the Cosmic Dawn Center of Excellence funded by the Danish National Research Foundation under the grant No. 140. 
\end{acknowledgements}


\bibliographystyle{aa} 
\bibliography{refs}

\end{document}